\documentclass[12pt,english]{article}
\usepackage[T1]{fontenc}
\usepackage[latin9]{inputenc}
\usepackage{graphicx}
\usepackage{esint}

\makeatletter
\newcommand{\lyxaddress}[1]{
\par {\raggedright #1
\vspace{1.4em}
\noindent\par}
}

\makeatother

\usepackage{babel}
\begin{document}

\title{Why Non-equilibrium is Different}

\author{J. R. Dorfman, T. R. Kirkpatrick, and J. V. Sengers}

\maketitle

\lyxaddress{{\large{}Institute for Physical Science and Technology, University
of Maryland, College Park, MD 20742, USA}}
\begin{abstract}
{\large{}The 1970 paper, ``Decay of the Velocity Correlation Function''
{[}Phys. Rev. ${\bf A1}$, 18 (1970), see also Phys. Rev. Lett. }\textbf{\large{}18}{\large{},
988, (1967){]} by Berni Alder and Tom Wainwright, demonstrated, by
means of computer simulations, that the velocity autocorrelation function
for a particle in a gas of hard disks decays algebraically in time
as $t^{-1},$ and as $t^{-3/2}$ for a gas of hard spheres. These
decays appear in non-equilibrium fluids and have no counterpart in
fluids in thermodynamic equilibrium. The work of Alder and Wainwright
stimulated theorists to find explanations for these ``long time tails''
using kinetic theory or a mesoscopic mode-coupling theory. This paper
has had a profound influence on our understanding of the non-equilibrium
properties of fluid systems. Here we discuss the kinetic origins of
the long time tails, the microscopic foundations of mode-coupling
theory, and the implications of these results for the physics of fluids.
We also mention applications of the long time tails and mode-coupling
theory to other, seemingly unrelated, fields of physics. We are honored
to dedicate this short review to Berni Alder on the occasion of his
90th birthday!}{\large \par}
\end{abstract}

\section{{\large{}Divergences in Non-equilibrium Virial Expansions}}

{\large{}N. N. Bogoliubov\cite{bogol}, by means of functional assumption
methods, and later M. S. Green\cite{msg1} and E. G. D. Cohen\cite{cohen1},
using cluster expansion methods, independently solved the outstanding
problem in the non-equilibrium statistical mechanics of gases at the
time, namely, to extend the Boltzmann transport equation to dense
gases as a power series expansion in the density of the gas. These
authors were able to formulate a generalized Boltzmann equation for
monatomic gases with short ranged central potentials, in the form
of a virial expansion of the collision operator whose successive terms
involved the dynamics of isolated groups of two, three, four,...,
particles interacting amongst themselves. That is the generalized
Boltzmann equation was written by these authors as\cite{cohen2,grnpic}
\begin{eqnarray}
\frac{\partial f({\bf r,{\bf v},t)}}{\partial t}+{\bf v\cdot\nabla_{{\bf r}}}f(({\bf r,{\bf v},t)} & = & J_{2}(f,f)+J_{3}(f,f,f)+\cdots.\label{eq:genbeq}
\end{eqnarray}
Here $f(({\bf r,{\bf v},t)}$ is the single particle distribution
function, for finding particles at position ${\bf r}$ with velocity
${\bf v}$ at time $t.$ The collision operator $J_{2}$ is the usual
Boltzmann, binary collision operator, while the $J_{j}$ are collision
operators determined by the dynamical events taking place among an
isolated group of $j$ particles. }{\large \par}

{\large{}At roughly the same time as the problem of generalizing the
Boltzmann equation to higher densities was being addressed, M. S.
Green\cite{msg2,msg4} and R. Kubo\cite{kubo} independently developed
a general method for expressing the transport coefficients appearing
in the linearized equations of fluid dynamics in terms of time integrals
of equilibrium time correlation functions of microscopic currents.
These expressions have the general form 
\begin{eqnarray}
\xi(n,T) & = & \int_{0}^{\infty}dt\left\langle j_{\xi}(0)j_{\xi}(t)\right\rangle _{eq}.\label{eq:GKform}
\end{eqnarray}
Here $\xi(n,T)$ is a transport coefficient such as the coefficient
of shear viscosity, thermal conductivity, }\emph{\large{}etc}{\large{}.
at fluid density $n$ and temperature, $T,$ the brackets denote an
equilibrium ensemble average, and $j_{\xi}(t)$ is the value of an
associated microscopic current at some time $t.$ An example that
will be important for our discussion is the case of tagged-particle
diffusion whereby one particle in a gas of mechanically identical
particles has some non-mechanical tag that enables one to follow its
diffusion in the gas. For this case, the diffusion coefficient $D$
is given by the Green-Kubo formula: 
\begin{eqnarray}
D(n,T) & = & \int_{0}^{\infty}dt\left\langle v_{x}(0)v_{x}(t)\right\rangle _{eq},\label{eq:slfdif}
\end{eqnarray}
where $v_{x}(t)$ is the $x-$component of the velocity of the tagged
particle}\footnote{{\large{}We mention that transport coefficients, characterizing non-equilibrium
flows are, in the Green-Kubo formalism, expressed in terms of time
correlation functions measured in an equilibrium ensemble. This is
consistent with Onsager's assumption that the final stages of the
relaxation of microscopic fluctuations about an equilibrium state
can be described by macroscopic hydrodynamic equations. Another example
occurs in the treatment of dynamic light scattering by fluids in equilibrium. }{\large \par}}{\large{}. }{\large \par}

{\large{}From the density expansion of the collision operator or by
an equivalent cluster expansion of the time correlation function expressions,
one can, at least in principle, obtain expressions for transport coefficients
of the gas as power series in the density, similar to the virial expansions
for the equilibrium properties of the same gas. This parallel development
indicated the existence of a ``super statistical mechanics''. whereby
both equilibrium and non-equilibrium properties of a gas can be expressed
in the form of virial, or power series, expansions in the density
of the gas, obtained by means of almost identical cluster expansion
methods. However it quickly became clear that this parallelism was
purely illusory, the non-equilibrium properties of a gas have almost
nothing in common with its equilibrium properties. The first indication
of this situation appeared in 1965 when Dorfman and Cohen\cite{dorcoh1},
among others\cite{brush}, discovered that almost every term in the
non-equilibrium virial expansions diverges! }{\large \par}

{\large{}The differences between the equilibrium and non-equilibrium
virial expansions have their origins in the type of correlations upon
which the virial coefficients depend. The equilibrium virial coefficients
depend only upon static correlations between a fixed number of particles
in contrast to the non-equilibrium virial coefficients which depend
mainly, if not exclusively, upon dynamical correlations produced by
sequences of collisions taking place between a fixed number of particles.
To be explicit, the equilibrium virial expansion for the pressure,
$p(n,T),$ of a gas at number density $n$ and at temperature $T,$
and the non-equilibrium virial expansion for the transport coefficient,
$\xi,$ of a gas at local number density $n$ and at local temperature
$T$ are given by 
\begin{eqnarray}
\frac{p(n,T)}{nk_{B}T} & = & 1+nb_{1}(T)+n^{2}b_{2}(T)+\cdots,\label{eq:pressvirex}\\
\frac{\xi(n,T)}{\xi_{0}(T)} & = & 1+n\sigma^{d}a_{1}^{(\xi)}(T)+(n\sigma^{d})^{2}a_{2}^{(\xi)}(T)+\cdots.\label{eq:diffvirexp}
\end{eqnarray}
with $k_{B}$ Boltzmann's constant. Here $\xi_{0}(T)$ is the low
density value of the transport coefficient as determined by the Boltzmann
equation for the gas and $\sigma$ is the range of the range of the
intermolecular force. The coefficients $b_{j-1}(T)$ are determined
by static correlations among $j$ interacting particles. The range
of these static correlations is at most $j\sigma.$ In contrast, the
non-equilibrium virial coefficient $a_{j-2}^{(\xi)}$ depends upon
correlated sequences of collisions taking place among a group of $j$
particles in infinite space and over an arbitrarily long time interval
between the first and final collision of the sequence. For the systems
under discussion here, all the equilibrium virial coefficients, $b_{j},$
are finite and of order $(\sigma^{d})^{j}$, where $d$ is the spatial
dimension of the system. However all but the first few non-equilibrium
virial coefficients diverge! For two-dimensional systems, the coefficients
$a_{1}^{(\xi)},a_{2}^{(\xi)},\ldots,$ all diverge\cite{dorco2}.
For three-dimensional systems, the coefficients $a_{2}$ and higher
all diverge. }{\large \par}

\begin{figure}[!h]

\centering{}\includegraphics[width=0.35\textwidth]{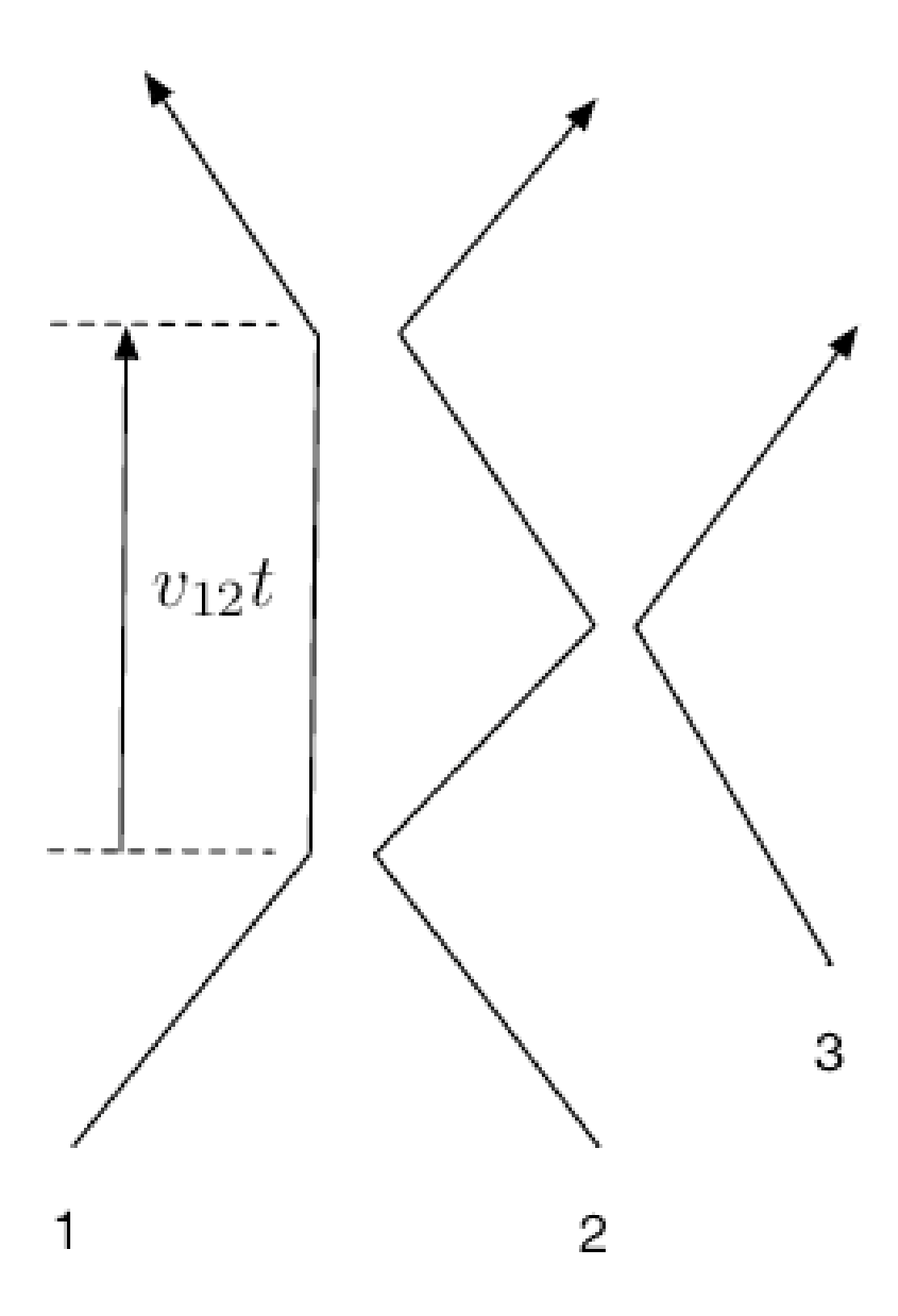}\includegraphics[width=0.35\textwidth]{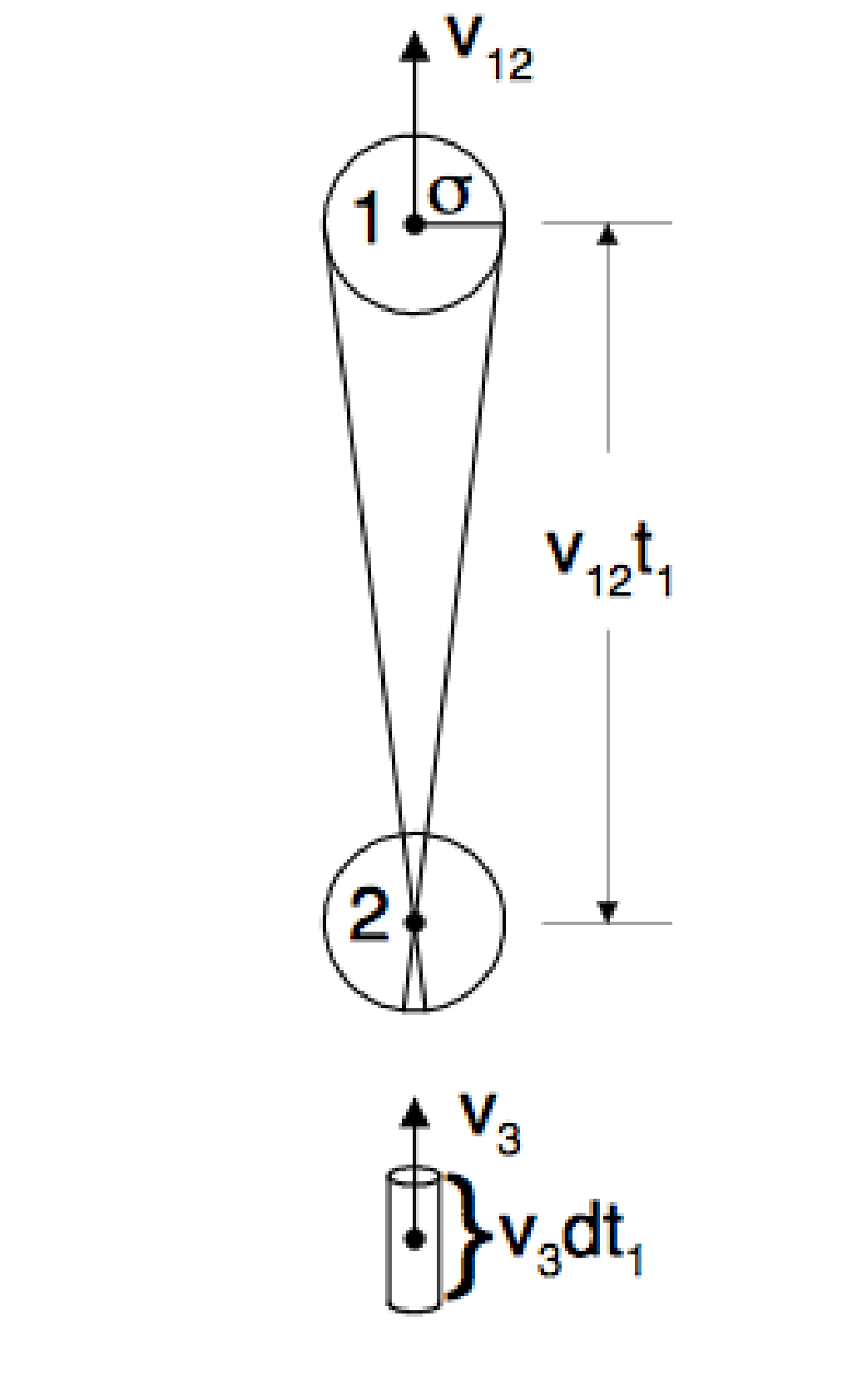}\protect\caption{Three particle recollision event: Particles $1$and $2$ collide at
some initial time, and due to an intermediate $(2,3)$ collision,
particles $1$ and $2$ collide at a time $t$ later. The second figure
is set in the rest frame of particle $2$, and shows the solid angle
into which particle $2$ must be scattered by particle $3$ to collide
with particle $1$ at time $t.$ Here $v_{12}$ is the relative velocity
of particles $1$ and $2$ after the first collision.}
\end{figure}

{\large{}The origin of these divergent coefficients can be easily
understood by considering the coefficient $a{}_{1}^{(\xi)},$ for
example. In Figure 1, we illustrate one of the three particle, correlated
collision sequences that contribute to this coefficient\cite{grnpic,dorco2}.
In this recollision event, particles $1$ and $2$ collide at some
initial instant, then later, particle $3$ collides with particle
$2$, in such a way that particles $1$ and $2$ collide again after
a time interval $t$ between the first and last collisions between
these two particles. The sequences take place in infinite space and
over arbitrarily large times, $t$. As illustrated in the Figure,
the dynamics is controlled by the solid angle into which particle
$2$ must be scattered when particle $3$ hits it. The phase space
region available for particle $3$ to cause the $(1,2)$ re-collision
between time $t$ and $t+dt$ is proportional to the solid angle and
is of the order $(\sigma/vt)^{d-1}dt$. The coefficient $a_{1}$ is
determined by the integration of this region over all possible time
intervals $t,$ and is clearly logarithmically divergent for $d=2.$
The coefficient $a_{1}^{(\xi)}$ is finite for $d=3,$ but the next
coefficient, $a_{2}^{(\xi)},$ is logarithmically divergent for three-dimensional
systems for similar reasons, and all higher coefficients diverge also,
as powers of the upper limit on the time integral which can be arbitrarily
large. Thus we can identify the essential difference between equilibrium
and non-equilibrium properties of gases: non-equilibrium processes
are due, among other things, to dynamical processes that can take
place over large spatial distances and over large times. These processes
cause long range and long time correlations among the particles in
the gas that are absent in equilibrium, except perhaps at critical
points, and even then, are of a qualitatively different origin. We
are now faced with another problem. The results of Bogoliubov, Green,
and Cohen are incomplete - their virial series are useless for descriptions
of processes that take place over times long compared to some microscopic
time due to the long time divergences in the terms in the virial series. }{\large \par}

\section{{\large{}The Ring Resummation}}

{\large{}It is clear what is causing the divergences in the non-equilibrium
expansions. A collective effect, mean free path damping of trajectories
has been ignored when deriving the virial expansions. We argued above
that the virial coefficients depend on the dynamics of isolated groups
of a fixed number of particles, and the time between any two collisions
in the troublesome correlated collision sequences can be arbitrarily
large. This is clearly unphysical. In a real gas, particles cannot
travel arbitrarily long distances between collisions without another
particle interrupting the motion of the particles by colliding with
one of them. That is to say, the typical distance between collisions
is a mean free path which in turn depends upon the gas density and
temperature. The probability of a particle moving a certain distance
is exponentially damped as the distance of travel becomes larger than
a few mean free path lengths. In essence, by insisting that the collision
operator or that the time correlation expressions be expanded in a
power series in the gas density, one has taken what should be an exponential
damping and expressed the exponential as a power series. Thus non-equilibrium
virial expansions are very misleading since they are the equivalent
of writing 
\begin{eqnarray}
e^{-nt} & = & 1-nt+\frac{1}{2}(nt)^{2}+\cdots,\label{eq:expon}
\end{eqnarray}
and trying to determine the behavior of the exponential by examining
individual terms on the right-hand side of its power series expansion.
It is clear that a more physical representation of the generalized
collision operator or of the time correlation function expressions
should be obtained by summing the most divergent terms in the virial
expansions and using the resummed expression, not the virial expansions.
This resummation was first carried out by K. Kawasaki and I. Oppenheim
in 1965\cite{kawopp}. They expressed the most divergent terms in
the virial expansions as ring events and were able to resum these
most divergent terms and to obtain expressions for transport coefficients
that should be well behaved, in contrast to the virial expansion representations}\footnote{{\large{}{}{}{}{}It is worth pointing out that nothing like this
has to be done for equilibrium virial expansion if the gas is composed
of particles interacting with short range forces. However, if the
gas is composed of particles interacting with infinite range Coulomb
potentials, a similar ring summation is necessary even in equilibrium.}{\large \par}}{\large{}. For three-dimensional systems one can use the resummed
expressions for transport coefficients to show that the logarithmic
divergence in the virial expansion is replaced by a logarithmic term
in the density, that results from including the mean free path damping
in the relevant collision integrals. Thus the first few terms in the
density expansion of the transport coefficients are 
\begin{equation}
\frac{\xi(n,T)}{\xi_{0}(T)}=1+a_{1}^{(\xi)}n\sigma^{3}+a_{2,ln}^{(\xi)}(n\sigma^{3})^{2}\ln n\sigma^{3}+a_{2,n}^{(\xi)}(n\sigma^{3})^{2}+\cdots.
\end{equation}
The coefficient of the linear term, $a_{1}^{(\xi)}$, had already
been calculated by Sengers\cite{sengers1} for hard spheres, based
on the analysis of this three-body collision integral by S. T. Choh
and G. E. Uhlenbeck\cite{chohuhl} and by Green\cite{msg3}. Also
for hard spheres the coefficient, $a_{2,ln}^{(\xi)},$ of the logarithmic
term has been calculated\cite{kamseng}, and estimates have been made
of the coefficient $a_{2,n}^{(\xi)}$ using the Enskog theory. In
Figure 2 we show the comparison of the theoretical and computer results
for the coefficients of self-diffusion, shear viscosity, and thermal
conductivity for a moderately dense gas of hard spheres\cite{dorsenkir}.
The agreement is quite good despite the fact that the coefficient
$a_{2,n}^{(\xi)}$ can only be estimated for reasons that will become
clear below}\footnote{{\large{}{}{}{}{}To anticipate this discussion, we mention that
the value of this coefficient depends upon the full time behavior
of the relevant time correlation functions or upon a good guess at
a lower cut-off of the time integrations.}{\large \par}}{\large{}. }
\begin{figure}[!h]
\begin{centering}
\includegraphics[angle=90,width=1\textwidth]{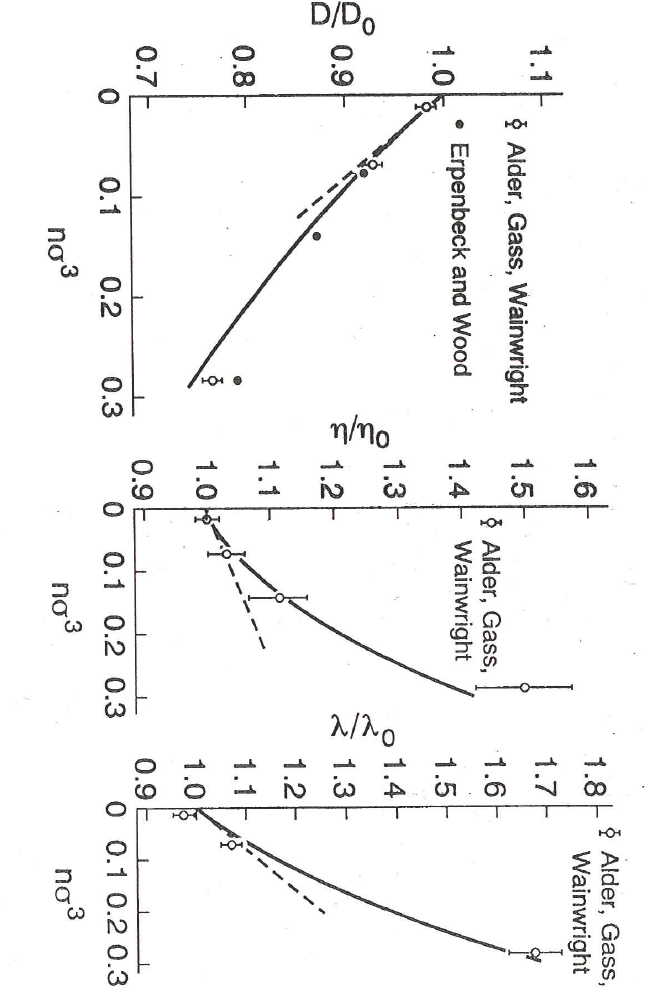}
\par\end{centering}

\begin{centering}
{\large{}\protect\caption{Comparison of the theoretical expressions, Eq. (7) for the transport
coefficients of self-diffusion, $D,$ the coefficient of shear viscosity,
$\eta$, and thermal conductivity, $\lambda,$ for a gas of hard spheres
with the results of molecular dynamics\cite{dorsenkir,agw,wwwerp2}.
The dashed lines correspond to keeping only the first two terms in
this expansion. The coefficient $a_{2,n}^{(\xi)}$ is estimated using
the Enskog theory.}
}
\par\end{centering}{\large \par}

\end{figure}
{\large \par}

{\large{}The non-analytic terms in a density expansion of a transport
coefficient have also been considered in quantum systems. Indeed,
very early on J. S. Langer and T. Neal\cite{langneal}, motivated
by the above classical work, pointed out that logarithmic terms appear
in the electrical conductivity in disordered electronic systems. It
can be argued that this sort of calculation, basically a quantum Lorentz
gas, is also relevant for the electron mobility, $\mu$, of excess
electrons in liquid helium. In this case the dimensionless density
expansion parameter, $\chi=4na_{s}^{2}\lambda$, involves the thermal
de Broglie wavelength, $\lambda=(2\pi^{2}\hbar^{2}/mk_{B}T)^{1/2}$,
the density of helium atoms, $n$, and the $s-$wave scattering length
$a_{s}$. Here $\hbar$ is Planck's constant and $m$ is the electron
mass. Wysokinski, Park, Belitz and Kirkpatrick\cite{Wysokinski_Park_Belitz_Kirkpatrick,wyospark}
have exactly computed $\mu$ up to and including terms of $O(\chi^{2})$
and obtained, 
\begin{equation}
\mu/\mu_{B}=1+\mu_{1}\chi+\mu_{2ln}\chi^{2}ln\chi+\mu_{2}\chi^{2}+o(\chi^{2}).\label{eq:dexp}
\end{equation}
Here $\mu_{B}$ is the Boltzmann equation value for $\mu$, and $\mu_{1}=-\pi^{3/2}/6,$
$\mu_{2ln}=(\pi^{2}-4)/32,$ and $\mu_{2}=0.236\ldots.$ Adams }\emph{\large{}et
al}{\large{} \cite{Adams_Browne_Paalanen} have concluded that existing
experiments give very good agreement with the value of the conductivity
given by Eq. (\ref{eq:dexp}).}{\large \par}

{\large{}To test experimentally the presence or absence of the logarithmic
term, Wysokinski }\emph{\large{}et al}{\large{} defined the function,
\begin{equation}
f(\chi)=[\mu/\mu_{B}-1-\mu_{1}\chi]/\chi^{2}.
\end{equation}
Theoretically, 
\begin{equation}
f(\chi)=\mu_{2ln}ln\chi+\mu_{2}\pm2\pi^{1/2}\chi,
\end{equation}
where the last term is an estimate of the $O(\chi^{3})$ contribution
to $\mu$. In Figure 3\cite{Wysokinski_Park_Belitz_Kirkpatrick,wyospark}
the theoretical prediction is shown for $0<\chi<0.7$ together with
experimental data\cite{schwartz}. The error bars shown assume a total
error of $3\%$ in $\mu/\mu_{B}$ and $4\%$ in $\chi$. To illustrate
the effect of the logarithmic term, the figure also shows what the
theoretical prediction would be if $\mu_{2ln}$ in Eq.(\ref{eq:dexp})
were zero.}
\begin{figure}[!h]

\begin{centering}
{\large{}\includegraphics[width=0.6\textwidth]{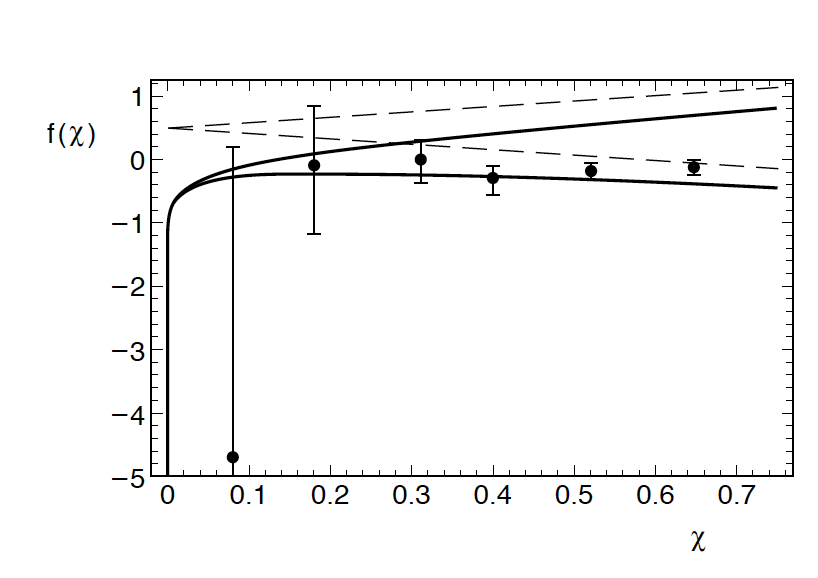}\protect\caption{The reduced mobility f, as defined in Eq. (9), \emph{vs}. the density
parameter $\chi$. The theoretical prediction of Wysokinski\emph{,
}Park, Belitz, and Kirkpatrick\emph{ }is for f to lie between the
two solid curves \cite{Wysokinski_Park_Belitz_Kirkpatrick}. The experimental
data are from Fig. 9 of Ref. \cite{schwartz} with error bars estimated
as in the two possible forms of Eq. (10). The broken lines show what
the theoretical prediction would be in the absence of the logarithmic
term in the density expansion.}
}
\par\end{centering}{\large \par}

\end{figure}
{\large \par}

{\large{}Following the discovery that logarithmic terms must appear
in non-equilibrium density expansions, there were strong indications
that something was still amiss in the kinetic theory for transport
coefficients. In 1966 R. Goldman\cite{goldman} argued that the resummed
expressions for transport coefficients contain time integrals of functions
with power-law decays. He identified the leading power as $t^{-3/2}$
for long times, for three-dimensional systems. In 1968 Y. Pomeau\cite{pomeau}
argued that for two-dimensional systems the Kawasaki- Oppenheim expressions
still diverge as time integrals of functions that decay as $t^{-1}$
for large times. }{\large \par}

{\large{}This was the situation just before the work of Alder and
Wainwright on the velocity auto-correlation function became known,
and before the appearance in 1970 of their paper in Physical Review
which stimulated so much work in non-equilibrium statistical mechanics,
and continues to reverberate even now with new and unexpected applications.}{\large \par}

\section{The Alder-Wainwright Paper of 1970: Long Time Tails}

{\large{}The papers by Alder and Wainwright in Physical Review Letters
in 1967\cite{aw1}, and most especially, that in Physical Review in
1970\cite{aw2} provided the spark that ignited the imaginations of
those of us working in kinetic theory. They considered gases of hard
spheres or of hard disks at moderate densities and by means of computer
simulated molecular dynamics, obtained the velocity correlation function
$\left\langle v_{x}(0)v_{x}(t)\right\rangle /\left\langle v_{x}^{2}\right\rangle $
for a range of times, scaled with the appropriate mean free time,
$t_{m},$ between collisions. Their results provided convincing evidence
that over a range of times, roughly $10\leq s=t/t_{m}\leq30,$ the
velocity autocorrelation functions decay as 
\begin{equation}
\frac{\left\langle v_{x}(0)v_{x}(s)\right\rangle _{eq}}{\left\langle v_{x}^{2}\right\rangle _{eq}}\simeq\alpha_{D}^{(d)}(n)s^{-d/2}.\label{eq:awltt}
\end{equation}
Here $\alpha_{D}^{(d)}(n)$ is a numerical coefficient that depends
on the density and the spatial dimension of the gas. The subscript
$D$ indicates that the time correlation function is the one needed
for the coefficient of self- or tagged particle diffusion through
Eq. (\ref{eq:slfdif}). Figure 4 shows their results for the three-dimensional
case.}{\large \par}

\begin{figure}[!h]

\begin{centering}
\includegraphics[width=0.6\textwidth]{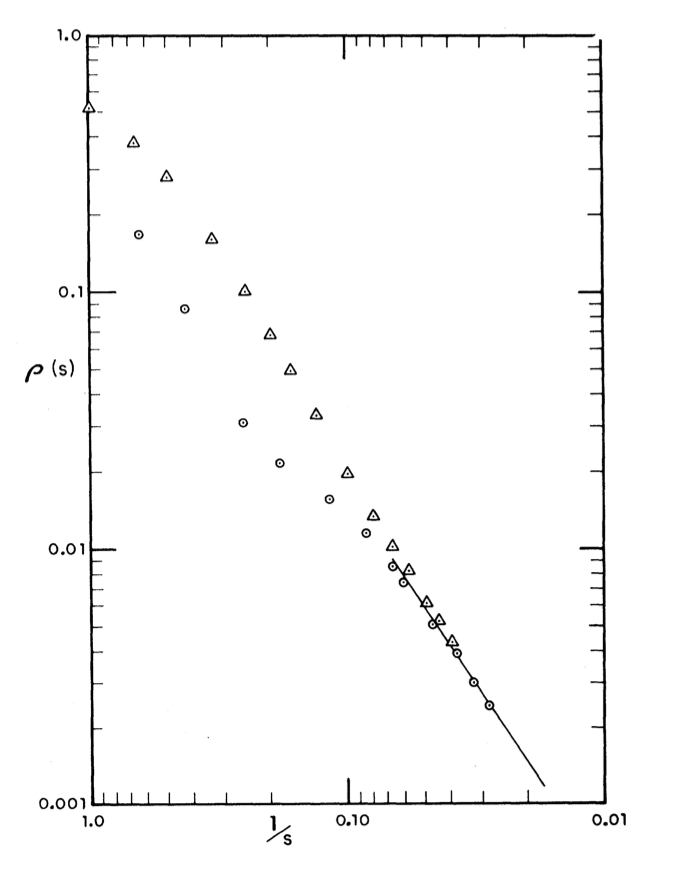}\protect\caption{The normalized velocity autocorrelation function as obtained by molecular
dynamics (triangles) for a gas of $500$ hard spheres\cite{aw2}.
The circles are results obtained using a hydrodynamical model developed
by Alder and Wainwright to explain their results.}

\par\end{centering}

\end{figure}

{\large{}Stimulated by these computer results and following theoretical
arguments of Goldman and Pomeau, Dorfman and Cohen\cite{dcltt1,dcltt2}
were able to show that these algebraic decays can be explained both
qualitatively and quantitatively by kinetic theory. They evaluated
the Kawasaki-Oppenheim ring summation, but in order to obtain results
appropriate for the densities studied by Alder and Wainwright, they
extended the summation result to higher densities by means of the
Enskog theory for dense hard ball gases\cite{dcltt3}. At the same
time Ernst, Hauge, and van Leeuwen\cite{ehvl} provided a mesoscopic
argument for these algebraic decays , or as they are called now, }\emph{\large{}long
time tails. }{\large{}The expression for the coefficient $\alpha_{D}^{(d)}(n)$
will serve to illustrate a general feature of the theoretical explanation
of the long time tails, 
\begin{equation}
\alpha_{D}^{(d)}(n)=c_{d}\left[(D+\nu)t_{m}\right]^{-d/2}.\label{eq:alphaD}
\end{equation}
Here $c_{d}$ is a numerical coefficient and proportional to $n^{2-d}$,
$D$ is the coefficient of self-diffusion, and $\nu=\eta/\rho$ is
the kinematic viscosity, $\eta$ is the coefficient of shear viscosity
and $\rho$ is the mass density of the fluid. The comparison of the
kinetic theory results, using the Enskog theory for the transport
coefficients with the results of Alder and Wainwright is illustrated
in Figure 5 \cite{dcltt1}. This provides conclusive proof that the
Alder-Wainwright results can be explained by kinetic theory when the
contributions of the most divergent terms are taken into account.}{\large \par}

\begin{figure}[!h]

\begin{centering}
\includegraphics[width=0.6\textwidth]{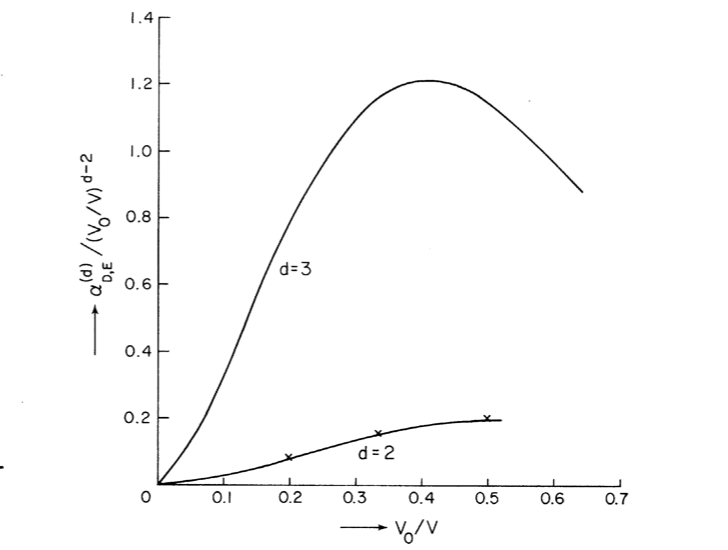}\protect\caption{The solid curves are the theoretical results obtained by Dorfman and
Cohen for the coefficients, $\alpha_{D,E}^{(d)}$ appearing in Eq.
(\ref{eq:alphaD}) for the velocity autocorrelation function, using
Enskog theory values for the transport coefficients\cite{dcltt1}.
The crosses represent values obtained from molecular dynamics by Alder
and Wainwright, for two-dimensional systems\cite{aw2}. Here $V/V_{0}$
is the ratio of the volume of the system to that at close packing
of the disks or spheres. }

\par\end{centering}

\end{figure}
{\large{}In Figure 6 we show later results of Wood and Erpenbeck\cite{wwwerp}
confirming those of Alder and Wainwright and they compared their results
with theoretical results including finite size effects. }{\large \par}

\begin{figure}[!h]

\begin{centering}
\includegraphics[width=0.6\textwidth]{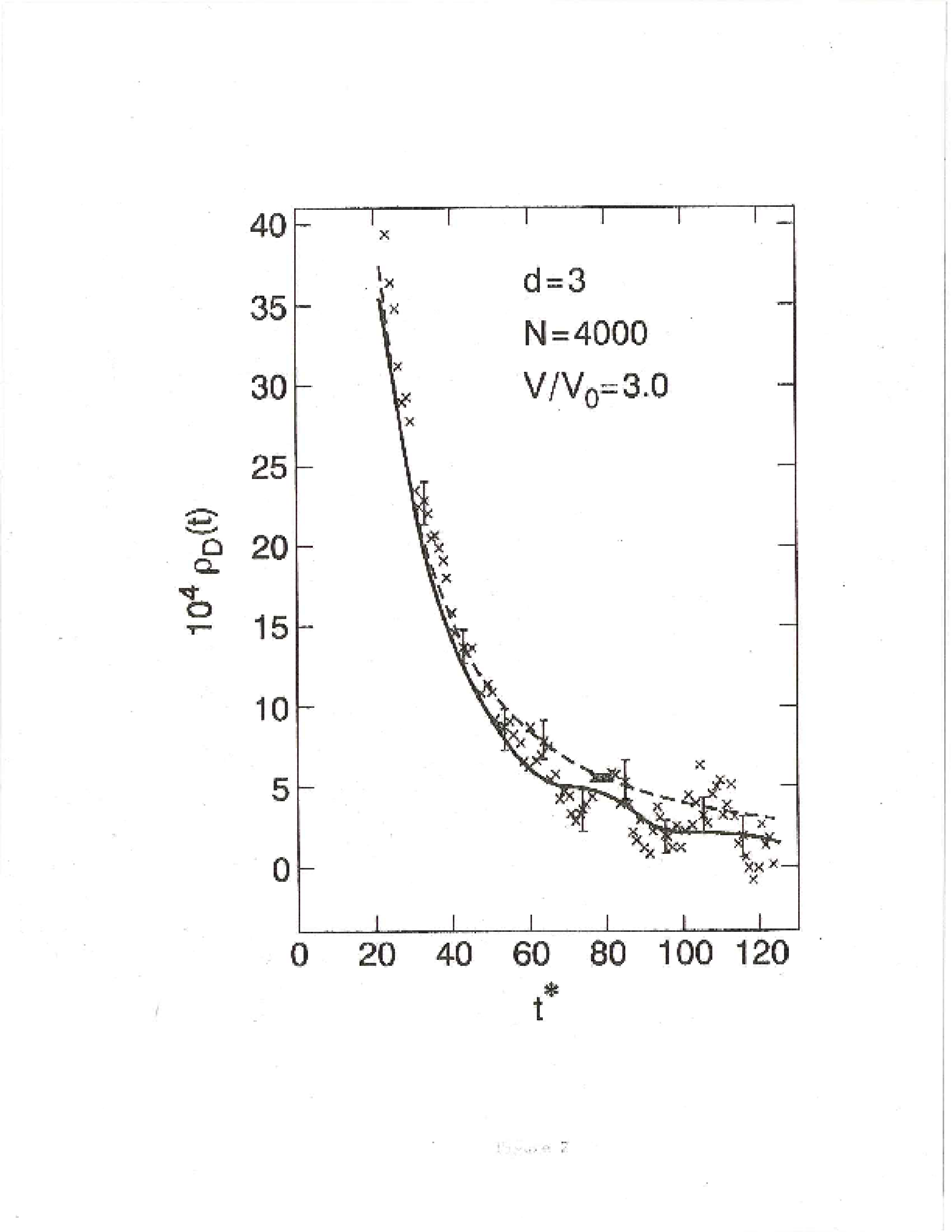}\protect\caption{Results of Wood and Erpenbeck for the velocity autocorrelation function
for a gas of $4000$ hard spheres at a volume of three times the close
packing volume\cite{wwwerp3}. Here $t^{*}$ is the time, measures
in mean free times. The crosses are computer results, the dashed line
is that given by Eq. (\ref{eq:awltt}), and the solid line represents
a complete evaluation of the mode coupling formula with all modes
taken into account and finite size corrections included.}

\par\end{centering}

\end{figure}

{\large{}It is important to note that $\alpha_{D}^{(d)}$ depends
upon the sum of two transport coefficients, in this case the coefficient
of self-diffusion and the kinematic viscosity. This is an indication
of the fact that the underlying microscopic processes generating the
tails are the coupling of microscopic hydrodynamic modes that exist
as fluctuations in fluids and are detected in dynamic light scattering
experiments}\footnote{{\large{}{}{}{}{}The Rayleigh and Brillouin peaks seen in dynamic
light scattering by an equilibrium fluid are due to microscopic heat
and sound modes appearing as fluctuations about equilibrium in the
fluid. }{\large \par}}{\large{}. The dynamical events taking place in the gas generate both
the modes and their coupling. A simple example will illustrate the
point. A somewhat oversimplified picture of a renormalized recollision
illustrated in Fig. 1 is shown in Figure 7 \cite{dorfanc}. Two particles
collide at some instant of time, then undergo an arbitrary number
of intermediate collisions before recolliding at time $t.$ One can
think of the motions of the two particles after their first collision
as random walks that cross at time $t.$ If we sit on one of the particles
we can imagine that the recollision is a random walk that returns
to the origin. A standard calculation in random walk theory shows
that the probability of a return to the origin after a time interval
$t$ is proportional to $(1/t)^{d/2}$. This time dependence is exactly
that of the long time tails, and the random walks represent hydrodynamic
processes such as diffusion that are coupled by the initial and final
collisions.}\footnote{{\large{}{}{}{}{}For tagged particle diffusion only one of the
initial colliding pair is followed, while the other particles in the
collision sequences can be any other particles in the fluid. The tagged
particle motion is represented by the appearance of the diffusion
coefficient in the long time tail result, Eq. (\ref{eq:alphaD}),
while the motions of the other particles in the sequence are represented
by the viscous mode contribution to this formula. }{\large \par}}

\begin{figure}[!h]
\begin{centering}
\includegraphics[width=0.45\textwidth]{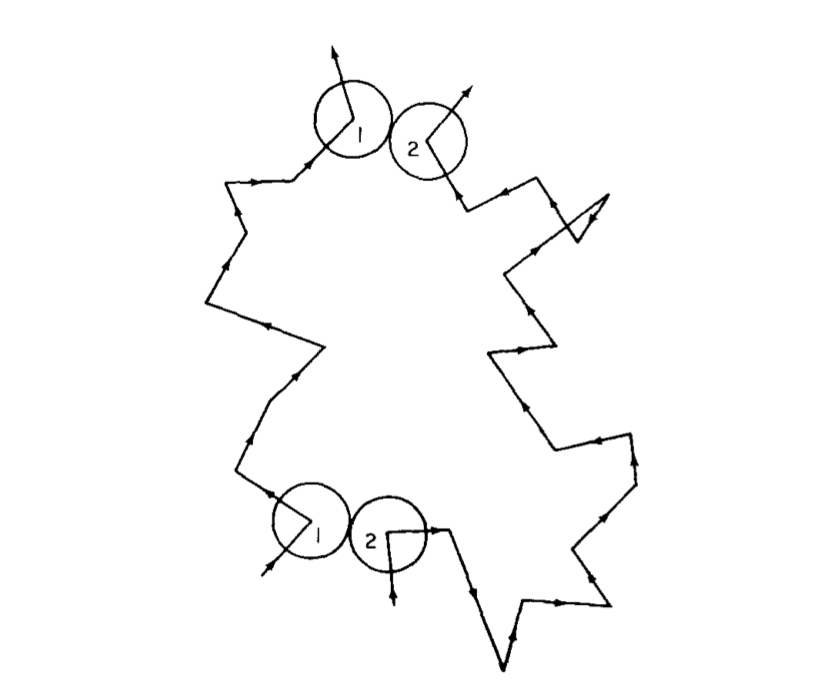}
\par\end{centering}

\protect\caption{A schematic version of a renormalized recollision sequence. Two particles
collide and then each of them undergoes a random walk produced by
collisions with other particles before they collide again after a
time $t.$}

\end{figure}

{\large{}We thus have, when this is all worked out properly, a microscopic
derivation of }\emph{\large{}mode-coupling theory}{\large{}, already
known from the work of L. P. Kadanoff and J. Swift\cite{kadsw} and
of Kawasaki\cite{kkaw} on the behavior of transport coefficients
near the critical point of a phase transition. In fact the Kadanoff-Swift
results are exactly the combined result of the long time tail processes
with the behavior of thermodynamic properties near a critical point.
We also mention that the transport coefficients appearing in the expression
for $\alpha_{D}^{(d)}$ have to be treated with some care. They cannot
be the full transport coefficients since those are determined by the
long time behavior of correlation functions. Instead, over the time
of the Alder-Wainwright studies these are to be seen as short time
contributions, thus accounting for the success of using the Enskog
expressions for the transport coefficients when comparing the computer
results with those from kinetic theory. }{\large \par}

\section{{\large{}Consequences of the Long Time Tails for Hydrodynamics}}

{\large{}The algebraic time decays of the time correlations and the
existence of generic long range correlations in non-equilibrium systems
have immediate consequences for microscopic derivations of the Navier
Stokes and higher-order hydrodynamic equations. The most immediate
of these is that the time correlation functions expressions for transport
coefficients diverge logarithmically with the upper limit of the time
integrals in the Green-Kubo formulas. For three-dimensional systems,
the Navier Stokes transport coefficients are finite but transport
coefficients in higher-order equations, such as the Burnett equations
diverge\cite{dcltt1,ehvl,erndor,pomres}. We are therefore faced with
the fact that our microscopic derivations of the fluid dynamics equations
have divergence problems. A number of studies have been carried out
in order to determine a more correct form of these equations, free
of divergence problems. The results are complicated and depend to
a certain extent on the transport process. For example, for two-dimensional
viscous flows, one finds that Newton's law of viscous friction must
be modified by the addition of non-linear logarithmic terms in the
velocity gradients. For three-dimensional systems, there is a non-analytic
correction to Newton's law. That is the off-diagonal terms of the
pressure tensor, $P_{xy},$ for example have the form\cite{erndorcic,onuki}
\begin{eqnarray}
P_{xy}^{(2)} & \simeq & -\tilde{\eta}X+aX\ln X+\cdots,\nonumber \\
P_{xy}^{(3)} & \simeq & -\eta X+bX|X|^{1/2}+\cdots,\nonumber \\
X & = & \frac{\partial u_{x}(y)}{\partial y}.\label{eq:shearflow}
\end{eqnarray}
Here $u_{x}(y)$ is the component of the fluid velocity, ${\bf u}$
that is a function of the coordinate in a perpendicular direction,
as is appropriate for shear flow. We see that for two-dimensional
systems, viscous flow is inherently nonlinear, since a coefficient
of shear viscosity defined by the limit $\lim_{X\to0}P_{xy}/X$ does
not exist. For three-dimensional systems the corrections to Newton's
law are non-analytic; in this case, a fractional power of the velocity
gradient appears. Physically, a finite shear rate weakens or makes
shorter range the correlations that cause the divergence problems.}{\large \par}

{\large{}The same considerations have also been applied to the case
of a stationary temperature gradient. Surprisingly, this case is very
different. A finite $\nabla T$ does not fix the divergence problem
in the two-dimensional heat conductivity, nor does it lead to non-analytic
terms in the three-dimensional heat flux\cite{kirkphd,kirkdoefnnoneq}.
This in turn implies that correlations in a non-equilibrium system
with a temperature gradient are of longer range, and more robust,
than a fluid with a velocity gradient. This observation is intimately
tied to the striking results discussed in the next section.}{\large \par}

{\large{}For three-dimensional systems the dispersion relation for
sound propagation in a gas also has a non-analytic form\cite{pomeau2,erndor},
\begin{equation}
\omega(k)=\pm ick+\frac{1}{2}\Gamma k^{2}+Ak^{5/2}+\cdots,\label{eq:sound}
\end{equation}
where $\omega(k)$ is the frequency of sound as a function of the
wave number, $k,$ $c$ is the velocity of sound and $\Gamma$ is
the sound damping constant. There are also an infinite number of terms
between $k^{2}$ and $k^{3}$, only the first of which is given here.
There is, indeed, some experimental evidence for the appearance of
the $k^{5/2}$ term in this dispersion relation as seen from neutron
scattering studies on liquid sodium\cite{morkel}. It is possible
to analyze the neutron scattering data in order to obtain values of
the frequency dependence of the Fourier transform, $Z(\omega)$ of
the velocity correlation function as a function of the frequency,
$\omega$. The long time tail in this function would then be seen
as a dependence of the Fourier transform on $\omega^{1/2}$. The results
of Morkel }\emph{\large{}et al}{\large{} are illustrated in Figure
8. The square root dependence is clear and the data are in good agreement
with the theory.}{\large \par}

\begin{figure}[!h]

\begin{centering}
\includegraphics[width=0.6\textwidth]{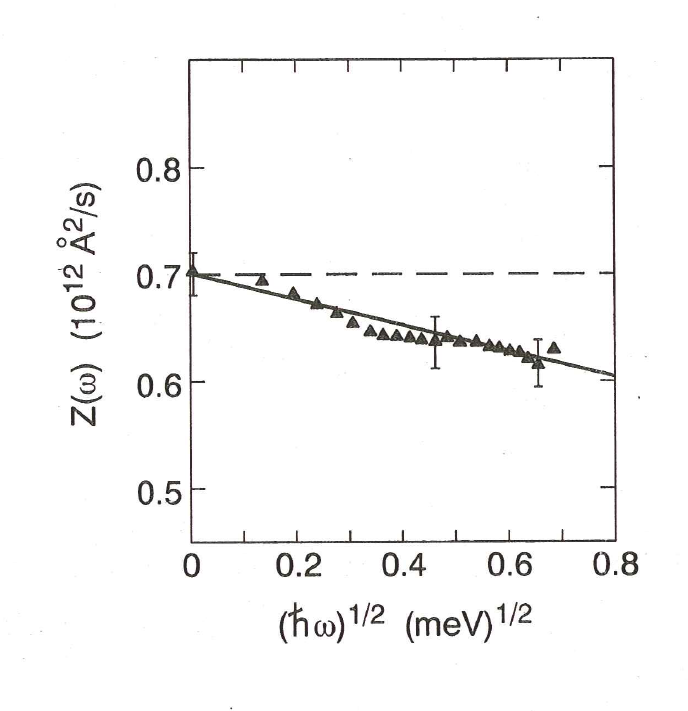}\protect\caption{The Fourier transform of the velocity autocorrelation function, $Z(\omega)$
as a function of the square root of the frequency, $\omega,$ for
atoms in liquid sodium as obtained from neutron scattering experiments
(triangles). The solid line is the theoretical result including mode-coupling
effects, while the dashed line omits them.\cite{morkel}}

\par\end{centering}

\end{figure}

{\large{}In general, very little is known about the complete structure
of the hydrodynamic equations, especially for two-dimensional systems.
Non-analytic terms, finite size effects, branch point structures,
and so on seem to be present. The only redeeming feature of all of
this is that these complications do not appreciably distort the results
obtained by using ordinary Navier Stokes hydrodynamics, even if, for
two dimensions we can only give approximate results for the transport
coefficients that appear in them. }{\large \par}

\section{Non-equilibrium Steady States}

{\large{}Very dramatic deviations from equilibrium behavior due to
mode-coupling effects causing long range spatial correlations can
be found in the properties of fluids maintained in non-equilibrium
stationary states. The first striking example of this difference was
discovered by Kirkpatrick\cite{kirkphd}, described in his doctoral
dissertation and in a subsequent series of papers by Kirkpatrick,
Cohen and Dorfman\cite{kcd1,kcd2,kcd3}. Confirmation of this work
was obtained by Sengers and co-workers in a series of light scattering
experiments on a fluid maintained in a steady state with a fixed temperature
gradient. As we noted above the structure factor for an equilibrium
fluid is, for small wave numbers, characterized by a central Rayleigh
peak and two Brillouin peaks on either side of the central peak. All
of this changes when a constant temperature gradient is imposed on
the system. Most dramatic of these effects is the enhancement of the
central peak by orders of magnitude, an enhancement due to the long
range spatial correlations in a non-equilibrium fluid. When a temperature
gradient is imposed on a fluid, the central peak of structure factor,
$S_{neq}(t,{\bf k})$ for a simple fluid, is given for small wave
numbers $k$ as a function of time, $t,$ by 
\begin{eqnarray}
S_{neq}(t,{\bf k}) & = & S_{0}\left[\left(1+A_{T}\right)e^{-D_{T}k^{2}t}-A_{\nu}e^{-\nu k^{2}t}\right],\nonumber \\
A_{T} & = & \frac{c_{P}}{T\left(\nu^{2}-D_{T}^{2}\right)}\left(\frac{\nu}{D_{T}}\right)\frac{\left(\hat{{\bf k}}_{\perp}\cdot\nabla T\right)^{2}}{k^{4}},\nonumber \\
A_{\nu} & = & \frac{c_{P}}{T\left(\nu^{2}-D_{T}^{2}\right)}\frac{\left(\hat{{\bf k}}_{\perp}\cdot\nabla T\right)^{2}}{k^{4}}.\label{eq:kfourted}
\end{eqnarray}
Here $S_{0}$ measures the intensity of the thermal fluctuations when
the fluid is in equilibrium, $c_{P}$ is the specific heat capacity
at constant pressure, $\nu,D_{T}$ are the coefficients of kinematic
viscosity and of thermal diffusivity, respectively, and $\hat{{\bf k}}_{\perp}$
is a unit vector in a direction perpendicular to that of the wave
vector, ${\bf k}$. It is important to note the inverse fourth power
of the wave number appearing in the coefficients $A_{T},A_{\nu}$,
and the proportionality to the square of the component of the temperature
gradient in a direction perpendicular to that of the wave vector.
The strong dependence on the wave number indicates quite clearly that
these effects are due to the long range nature of the spatial correlations
in the fluid, and these terms vanish for zero temperature gradient.
All the thermodynamic and transport coefficients are known for toluene,
for example, so that a direct comparison of theory and experiment
can be carried out, as was done by Sengers and co-workers\cite{segseng}}\emph{\large{}.
}{\large{}The results are given in Figure 9. The agreement of theory
and experiment is excellent. }{\large \par}

\begin{figure}[!h]

\begin{centering}
\includegraphics[width=0.6\textwidth]{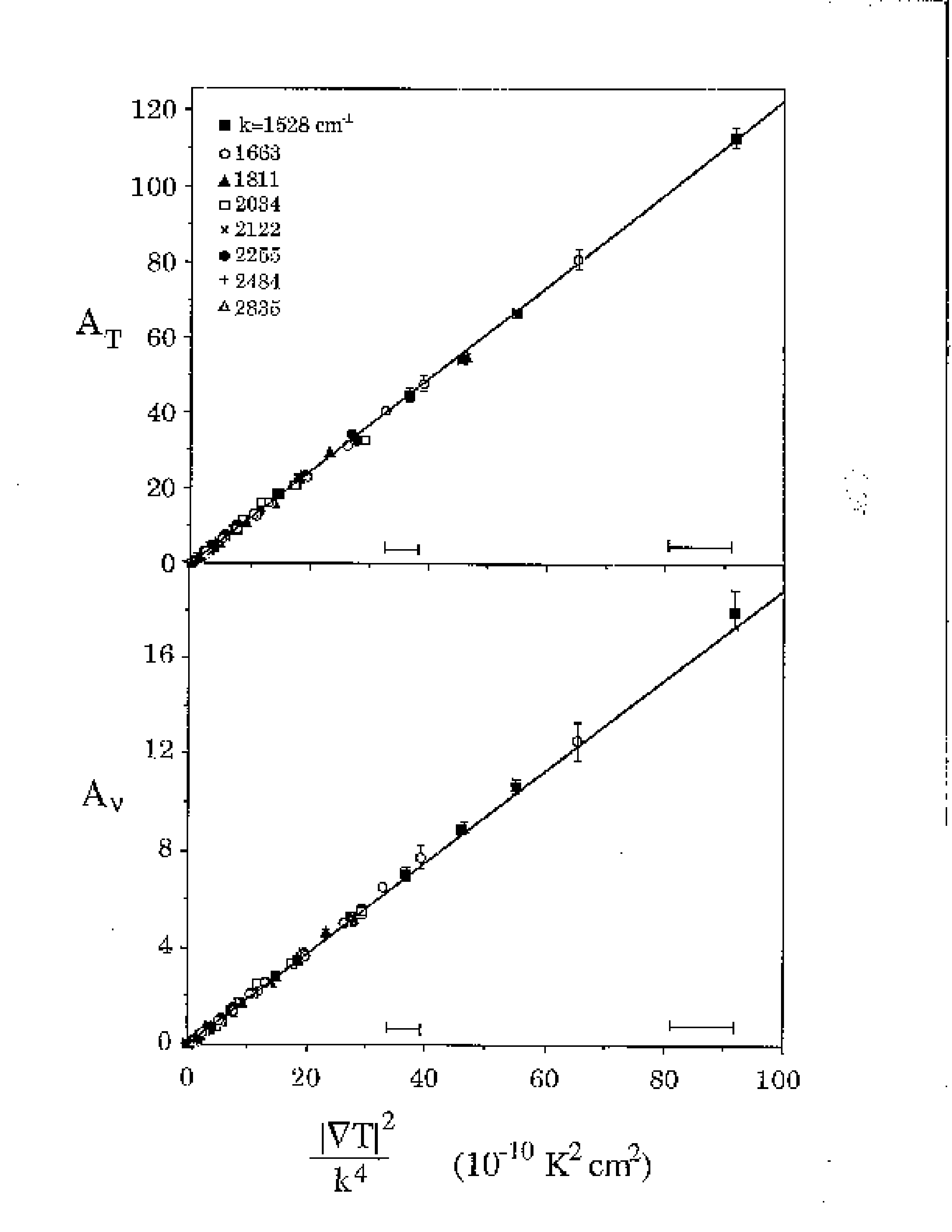}\protect\caption{The coefficients $A_{T}$ and $A_{\nu}$ as a function of wave number
as measured in light scattering experiments by Sengers \emph{et al}
\cite{segseng}\emph{.} The solid lines are theoretical values with
no adjustable parameters.}

\par\end{centering}

\end{figure}

{\large{}Generally, long-ranged fluctuations will also induce a so-called
Casimir force in a confined fluid\cite{kard}. A well known example
is the Casimir effect due to critical fluctuations in equilibrium
fluids\cite{fisdg,krech}. Critical fluctuations roughly vary as $k^{-2}$,
while the above non-equilibrium fluctuations vary as $k^{-4}$. Hence,
as shown by Kirkpatrick, Ortiz de Zárate, and Sengers\cite{kos1,kos2},
Casimir effects in confined non-equilibrium fluids that are substantially
larger than critical Casimir effects in equilibrium fluids. As an
example, we consider a liquid between two horizontal thermally conducting
plates separated by a distance $L$ and subject to a stationary temperature
gradient $\nabla T$. The non-equilibrium Casimir effects are two
fold. First, there will be a fluctuation-induced non-equilibrium contribution
to the density profile as a function of height. Second, the fluctuations
cause an additional non-equilibrium pressure contribution, $\bar{p}_{NE},$
to the equilibrium pressure such that\cite{kos1,kos2}}{\large \par}

{\large{}
\begin{equation}
\overline{p}_{{\rm NE}}=\frac{c_{p}k_{{\rm B}}\overline{T_{0}}^{2}\left(\gamma-1\right)}{96\pi D_{T}\left(\nu+D_{T}\right)}\widetilde{B}F_{0}L\left(\frac{\nabla T_{0}}{\overline{T_{0}}}\right)^{2},\label{eq:avp}
\end{equation}
with, 
\begin{equation}
\widetilde{B}=\left[1-\frac{1}{\alpha c_{p}}\left(\frac{\partial c_{p}}{\partial T}\right)_{p}+\frac{1}{\alpha^{2}}\left(\frac{\partial\alpha}{\partial T}\right)_{p}\right],\label{eq:B}
\end{equation}
Here $\alpha$ is the thermal expansion coefficient, and $\gamma$
is the ratio of the isobaric and isochoric heat capacities. The coefficient
$F_{0}$ is a numerical constant whose value depends on the boundary
conditions for the velocity fluctuations that are coupled with the
temperature fluctuations through the temperature gradient. For stress-free
boundary conditions $F_{0}=1$. Just as in Eq. (\ref{eq:kfourted}),
all thermophysical properties, including the temperature $T$, can
to a good approximation be identified with their average values in
the liquid layer. Note that, for a given value of the temperature
gradient $\nabla T$, the fluctuation-induced pressure increases with
$L$. The physical reason is that the dependence of the fluctuations
implies that in real space the correlations scale with the system
size. This non-equilibrium pressure contribution corresponds to a
nonlinear Onsager-like cross effect\cite{kos1,kos2}:}{\large \par}

{\large{}
\begin{equation}
\bar{p}_{NE}=\kappa_{NL}(\nabla T)^{2},\label{burnett1}
\end{equation}
where $\kappa_{NL}$ is a coefficient in the Burnett equations mentioned
earlier in Section 4. Comparing Eqs. (\ref{eq:avp}) and (\ref{burnett1}),
we see that the non-equilibrium fluctuation-induced pressure is directly
related to the divergence of the nonlinear Burnett coefficient $\kappa_{NL}$
with increasing $L$. Experimentally, it may be more convenient to
investigate the fluctuation-induced pressure as a function of the
temperature difference $\delta T=L\nabla T$. Then 
\begin{equation}
\bar{p}_{NE}\propto\frac{1}{L}\left(\frac{\delta T}{T}\right)^{2}.\label{eq:burnett2}
\end{equation}
This result may be compared with the critical Casimir pressure in
equilibrium fluids: 
\begin{equation}
p_{c}=\frac{k_{B}T}{L^{3}}\Theta\left(\frac{L}{\xi}\right),\label{eq:critp}
\end{equation}
where $\xi$ is the correlation length. From Eq. (\ref{eq:critp}),
we have estimated that for water at $298$ K in a layer with $1$
micron and with $\delta T=25$ K , will be of the order of a Pa, while
$p_{c}$ is of the order of a milli-Pa for the same distance. Actually,
at $L=1$ mm, $\bar{p}_{NE}$ becomes already of the same order of
magnitude as $p_{c}$ at $L=1$ micron. One should also note that
the critical Casimir effect can only be observed in fluids near a
critical point, while the non-equilibrium Casimir effect will be generically
present in liquids at any temperature and density. We conclude that
thermal fluctuations in non-equilibrium fluids are fundamentally different
from thermal fluctuations in equilibrium fluids. }{\large \par}

\section{Long Time Tail Phenomena in Other Contexts}

{\large{}It is quite remarkable how often one encounters situations
in other physical contexts where the long time tails or, equivalently,
mode-coupling theory play an important role. Here we list just a few
examples.}{\large \par}

\subsection*{Critical Phenomena}

{\large{}As we mentioned earlier, mode-coupling theory was developed
more or less intuitively by Kadanoff and Swift in order to explain
the behavior of transport coefficients near the critical points of
phase transitions, such as the liquid-gas transition. It was known
from experiments of Sengers, carried out in the 1960's, that the coefficient
of thermal conductivity diverges near this critical point as recently
reviewed by Anisimov\cite{anis}. In this situation both static and
dynamic correlations have a long range. Later experiments of Sengers
and co-workers confirmed both effects\cite{chang,burst}. The results
were a combination of long time tail effects underlying mode-coupling
theory with the effects of the singular behavior of thermodynamic
properties of the fluid near its critical point. Other and related
applications of mode-coupling theory to the liquid-glass transition
have been important for the theory of glasses but we will not comment
on that work here. }{\large \par}

\subsection*{Weak Localization}

{\large{}In Section 2 we mentioned that in the mid $1960's$ Langer
and Neal\cite{langneal} showed that a logarithmic term appears in
the conductivity in disordered electron systems. It wasn't until the
late $1970's$ that the dynamical consequences of the correlations
that lead to the logarithmic term, basically quantum long-time tail
effects, were studied and understood\cite{Gorkov_Larkin_Khmelnitskii,Abraham_Anderson_Licciardello_Ramakrishnan}}\footnote{{\large{}A review article that stresses the generality of long time
tail phenomena in the context of a variety of closely related phenomena,
both classical and quantum, is given in Ref. \cite{Belitz_Kirkpatrick_Vojta}.}{\large \par}}{\large{}. This opened up the field of what became known as weak localization
in condensed matter physics which in turn is closely connected to
the phenomenon of Anderson localization\cite{Anderson}. Among other
things, the ultimate conclusion was that the effects were so strong
that at zero temperature a two-dimensional system is always an insulator\cite{Wegner_1979,Abraham_Anderson_Licciardello_Ramakrishnan,Schafer_Wegner,Wegner_1980}.
At finite temperature there are logarithmic temperature non-analyticities
that decrease the conductivity as $T$ is lowered. In three dimensions
there are weaker, but still important non-analyticities in both temperature
and frequency. All of these effects have been measured in great detail.
For reviews see Refs. \cite{Lee_Ramakrishnan,Belitz_Kirkpatrick_1994}.}{\large \par}

\subsection*{Cosmology}

{\large{}There appears to be a deep and interesting connection between
long time tail phenomena that we have been discussing here and the
results of investigations of the dynamics of black hole horizons.
The cosmology community has become aware of the results of non-equilibrium
statistical mechanics, in particular, the existence of long time tails
and their anomalous effects on the equations of fluid dynamics. We
will not go into the details but it is worth mention the titles of
a few recent papers: ``Hydrodynamic Long Time Tails from Anti de
Sitter Space'' by S. Caron-Huot and O. Saremi\cite{saremi}, and
``Hydrodynamic Fluctuations, Long Time Tails, and Supersymmetry''
by P. Kovtun and L. G. Yaffe\cite{kovtyaf}, among others. Such connections
reinforce the notion gained from experience that across a wide swath
of physics, people, perhaps without being aware of it, are working
on the same or closely related problems, and the only difference is
in the mathematical language used to describe them.}{\large \par}

\section{Conclusion}

{\large{}The paper has given a brief review of the history of kinetic
theory and related non-equilibrium statistical mechanics with an emphasis
of the work of Alder and Wainwright as described in their 1970 paper.
Alder and Wainwright helped consolidate prior work in kinetic theory
and stimulated much more work in theoretical, experimental, and computational
physics. We hope that we have made clear the profound influence the
1970 paper has had on non-equilibrium statistical mechanics and on
fields that on first sight might seem to be distantly related but
on closer inspection turn out to be closely related after all. We
are pleased to dedicate this paper to our friend, colleague and mentor,
Berni Alder, on the occasion of his 90th birthday!}{\large \par}

\section{Acknowledgement}

{\large{}The authors would like to thank D. Belitz and D. Thirumalai
for helpful discussions and Y. Bar Lev and A. Nava-Tudela for their
considerable help with the preparation of this paper. They would also
like to thank E. G. D. Cohen for useful and productive conversations
over a period of many years. TRK would like to thank the NSF for support
under Grant No. DMR-1401449}{\large \par}

\bibliographystyle{unsrt}
%\bibliography{alderbib1201}

\begin{thebibliography}{10}

\bibitem{bogol}
N.~N. Bogoliubov.
\newblock In J.~de~Boer and G.~E. Uhlenbeck, editors, {\em Studies in
  Statistical Mechanics}, volume~1, pages 1--118. North-Holland, Amsterdam,
  1961.

\bibitem{msg1}
M.~S. Green.
\newblock {\em J. Chem. Phys.}, 25:836, 1956.

\bibitem{cohen1}
E.~G.~D. Cohen.
\newblock {\em Physica}, 28:1025, 1962.

\bibitem{cohen2}
E.~G.~D. Cohen.
\newblock {\em J. Math. Phys.}, 4:183, 1963.

\bibitem{grnpic}
M.~S. Green and R.~A. Piccirelli.
\newblock {\em Phys. Rev.}, 132:1388, 1963.

\bibitem{msg2}
M.~S. Green.
\newblock {\em J. Chem. Phys.}, 20:1281, 1952.

\bibitem{msg4}
M.~S. Green.
\newblock {\em J. Chem. Phys.}, 22:398, 1954.

\bibitem{kubo}
R.~Kubo.
\newblock {\em J. Phys. Soc. Japan}, 12:370, 1957.

\bibitem{dorcoh1}
J.~R. Dorfman and E.~G.~D. Cohen.
\newblock {\em Phys. Lett.}, 16:124, 1965.

\bibitem{brush}
S.~G. Brush.
\newblock {\em Kinetic Theory}, volume~3.
\newblock Pergamon, New York, 1972.

\bibitem{dorco2}
J.~R. Dorfman and E.~G.~D. Cohen.
\newblock {\em J. Math. Phys.}, 8:282, 1967.

\bibitem{kawopp}
K.~Kawasaki and I.~Oppenheim.
\newblock {\em Phys. Rev.}, 139:1763, 1965.

\bibitem{sengers1}
J.~V. Sengers.
\newblock In W.~E. Brittin, editor, {\em Boulder Lectures in Theoretical
  Physics}, volume IX C, pages 335--374. Gordon and Breach, 1967.

\bibitem{chohuhl}
S.~T. Choh and G.~E. Uhlenbeck.
\newblock The kinetic theory of dense gases.
\newblock Technical report, University of Michigan, 1958.

\bibitem{msg3}
M.~S. Green.
\newblock {\em Phys. Rev.}, 136:905, 1964.

\bibitem{kamseng}
B.~Kamgar-Parsi and J.~V. Sengers.
\newblock {\em Phys. Rev. Lett.}, 51:2163, 1983.

\bibitem{dorsenkir}
J.~R. Dorfman, T.~R. Kirkpatrick, and J.~V. Sengers.
\newblock In {\em Ann. Rev. Phys. Chem.}, volume~45, page 213. Annual Reviews,
  1994.

\bibitem{agw}
B.~J. Alder, D.M. Gass, and T.~E. Wainwright.
\newblock {\em J. Chem. Phys.}, 53:3813, 1970.

\bibitem{wwwerp2}
J.~J. Erpenbeck and W.~W. Wood.
\newblock {\em Phys. Rev. A}, 43:4254, 1991.

\bibitem{langneal}
J.~S. Langer and T.~Neal.
\newblock {\em Phys. Rev. Lett.}, 16:984, 1966.

\bibitem{Wysokinski_Park_Belitz_Kirkpatrick}
K.~I. Wysokinski, W.~Park, D.~Belitz, and T.~R. Kirkpatrick.
\newblock {\em Phys. Rev. Lett.}, 73:2571, 1994.

\bibitem{wyospark}
K.~I. Wysokinski, W.~Park, D.~Belitz, and T.~R. Kirkpatrick.
\newblock {\em Phys. Rev. E}, 52:612, 1995.

\bibitem{Adams_Browne_Paalanen}
P.~W. Adams, D.~Browne, and M.~A. Paalanen.
\newblock {\em Phys. Rev. B}, 45:8837, 1992.

\bibitem{schwartz}
K.~Schwartz.
\newblock {\em Phys. Rev. B}, 21:5125, 1980.

\bibitem{goldman}
R.~Goldman.
\newblock {\em Phys. Rev. Lett.}, 17:130, 1966.

\bibitem{pomeau}
Y.~Pomeau.
\newblock {\em Phys. Rev. A}, 3:1174, 1971.

\bibitem{aw1}
B.~J. Alder and T.~E. Wainwright.
\newblock {\em Phys. Rev. Lett.}, 18:988, 1967.

\bibitem{aw2}
B.~J. Alder and T.~E. Wainwright.
\newblock {\em Phys. Rev. A}, 1:18, 1970.

\bibitem{dcltt1}
J.~R. Dorfman and E.~G.~D. Cohen.
\newblock {\em Phys. Rev. Lett.}, 25:1257, 1970.

\bibitem{dcltt2}
J.~R. Dorfman and E.~G.~D. Cohen.
\newblock {\em Phys. Rev. A}, 6:776, 1972.

\bibitem{dcltt3}
J.~R. Dorfman and E.~G.~D. Cohen.
\newblock {\em Phys. Rev. A}, 12:292, 1975.

\bibitem{ehvl}
M.~H. Ernst, E.~H. Hauge, and J.~M.~J. van Leeuwen.
\newblock {\em Phys. Rev. Lett.}, 25:1254, 1970.

\bibitem{wwwerp}
W.~W. Wood and J.~J. Erpenbeck.
\newblock In {\em Ann. Rev. Phys. Chem.}, volume~27, page 319. Annual Reviews,
  1976.

\bibitem{wwwerp3}
W.~W. Wood and J.~J. Erpenbeck.
\newblock {\em Ann. Rev. Phys. Chem.}, 27:319, 1976.

\bibitem{dorfanc}
J.~R. Dorfman.
\newblock {\em Physica A}, 106:77, 1981.

\bibitem{kadsw}
L.~P. Kadanoff and J.~Swift.
\newblock {\em Phys. Rev.}, 166:89, 1966.

\bibitem{kkaw}
K.~Kawasaki.
\newblock {\em Ann. Phys.}, 61:1, 1970.

\bibitem{erndor}
M.~H. Ernst and J.~R. Dorfman.
\newblock {\em J. Stat. Phys.}, 12:311, 1975.

\bibitem{pomres}
Y.~Pomeau and P.~Resibois.
\newblock {\em Phys. Rept.}, 19:63, 1975.

\bibitem{erndorcic}
M.~H. Ernst, B.~Cichocki, J.~R. Dorfman, J.~Sharma, and H.~van Beijeren.
\newblock {\em J. Stat. Phys.}, 18:237, 1978.

\bibitem{onuki}
A.~Onuki.
\newblock {\em Phys. Lett. A}, 70:31, 1979.

\bibitem{kirkphd}
T.~R. Kirkpatrick.
\newblock PhD thesis, Rockefeller University, New York, 1981.

\bibitem{kirkdoefnnoneq}
T.~R. Kirkpatrick and J.~R. Dorfman.
\newblock {\em Phys. Rev. E}, 92:022109, 2015.

\bibitem{pomeau2}
Y.~Pomeau.
\newblock {\em Phys. Rev. A}, 7:1134, 1973.

\bibitem{morkel}
C.~Morkel and C.~Gronemeyer.
\newblock {\em Z. Phys. B}, 72:433, 1988.

\bibitem{kcd1}
T.~R. Kirkpatrick, E.~G.~D. Cohen, and J.~R. Dorfman.
\newblock {\em Phys. Rev. A}, 26:950, 1982.

\bibitem{kcd2}
T.~R. Kirkpatrick, E.~G.~D. Cohen, and J.~R. Dorfman.
\newblock {\em Phys. Rev. A}, 26:972, 1982.

\bibitem{kcd3}
T.~R. Kirkpatrick, E.~G.~D. Cohen, and J.~R. Dorfman.
\newblock {\em Phys. Rev. A}, 26:995, 1982.

\bibitem{segseng}
P.~N. Segr{\'e}, R.~W. Gammon, J.~V. Sengers, and B.~Law.
\newblock {\em Phys. Rev. A}, 45:714, 1992.

\bibitem{kard}
M.~Kardar and R.~Golestanian.
\newblock {\em Rev. Mod. Phys.}, 71:1233, 1999.

\bibitem{fisdg}
M.~E. Fisher and P.~de~Gennes.
\newblock {\em C. R. Acad. Sci. Paris B}, 287:207, 1978.

\bibitem{krech}
M.~Krech.
\newblock {\em The Casimir Effect in Critical Systems}.
\newblock World Scientific, Singapore, 1994.

\bibitem{kos1}
T.~R. Kirkpatrick, J.~M.~Ortiz de~Z{\'a}rate, and J.~V. Sengers.
\newblock {\em Phys. Rev. Lett.}, 110:235902, 2013.

\bibitem{kos2}
T.~R. Kirkpatrick, J.~M.~Ortiz de~Z{\'a}rate, and J.~V. Sengers.
\newblock {\em Phys. Rev. E}, 89:022145, 2014.

\bibitem{anis}
M.~A. Anisimov.
\newblock {\em Int. J. Thermophys.}, 32:2001, 2011.

\bibitem{chang}
R.~F. Chang, H.~Burstyn, J.~V. Sengers, and A.~J. Bray.
\newblock {\em Phys. Rev. Lett.}, 37:1481, 1976.

\bibitem{burst}
H.~C. Burstyn and R.~F. Chang.
\newblock {\em Phys. Rev. Lett.}, 44:410, 1980.

\bibitem{Gorkov_Larkin_Khmelnitskii}
L.~P. Gorkov, A.~Larkin, and D.~E. Khmelnitskii.
\newblock {\em JETP Lett.}, 30(228), 1979.

\bibitem{Abraham_Anderson_Licciardello_Ramakrishnan}
E.~Abrahams, P.~W. Anderson, D.~C. Licciardello, and T.~V. Ramakrishnan.
\newblock {\em Phys. Rev. Lett}, 42:673, 1979.

\bibitem{Belitz_Kirkpatrick_Vojta}
D.~Belitz, T.~R. Kirkpatrick, and T.~Vojta.
\newblock {\em Rev. Mod. Phys.}, 77:579, 2005.

\bibitem{Anderson}
P.~W. Anderson.
\newblock {\em Phys. Rev.}, (1492), 1958.

\bibitem{Wegner_1979}
F.~Wegner.
\newblock {\em Z. Phys. B}, 35:207, 1979.

\bibitem{Schafer_Wegner}
L.~Schafer and F.~Wegner.
\newblock {\em Z. Phys. B}, 231:113, 1980.

\bibitem{Wegner_1980}
F.~Wegner.
\newblock {\em Z. Phys. B}, 36:209, 1980.

\bibitem{Lee_Ramakrishnan}
P.~A. Lee and T.~V. Ramakrishnan.
\newblock {\em Rev. Mod. Phys.}, 57:287, 1985.

\bibitem{Belitz_Kirkpatrick_1994}
D.~Belitz and T.~R. Kirkpatrick.
\newblock {\em Rev. Mod. Phys.}, 66:261, 1994.

\bibitem{saremi}
S.~Caron-Huot and O.~Saremi.
\newblock {\em J. High Ener. Phys.}, 13:1, 2010.

\bibitem{kovtyaf}
P.~Kovtun and L.~G. Yaffe.
\newblock {\em Phys. Rev. D}, 68:025007, 2003.

\end{thebibliography}

\end{document}